\documentstyle[prl,twocolumn,aps]{revtex}
\input psfig
\begin{document}
\draft

 \newcommand{\mytitle}[1]{
 \twocolumn[\hsize\textwidth\columnwidth\hsize
 \csname@twocolumnfalse\endcsname #1 \vspace{1mm}]}

\mytitle{
\title{Real-Time-RG Analysis of the Dynamics of the Spin-Boson Model}

\author{Markus Keil$^1$ and Herbert Schoeller $^{2}$}

\address{
$^1$ Institut f\"ur Theoretische Physik,
Universit\"at G\"ottingen, 37073 G\"ottingen, Germany\\
$^2$Forschungszentrum Karlsruhe, Institut f\"ur 
Nanotechnologie, 76021 Karlsruhe, Germany\\
}

\date{\today}

\maketitle

\begin{abstract}
Using a real-time renormalization group method we determine the
complete dynamics of the spin-boson model with ohmic dissipation for
coupling strengths $\alpha\lesssim 0.1-0.2$. We calculate the relaxation
and dephasing time, the static susceptibility and correlation
functions. Our results are consistent with quantum Monte Carlo
simulations and the Shiba relation.
We present for the first time reliable results for finite cutoff
and finite bias in a regime where perturbation theory
in $\alpha$ or in tunneling breaks down. Furthermore, an unambigious
comparism to results from the Kondo model is achieved.
\end{abstract}
}

{\it Introduction}.
The spin-boson model is one of the most fundamental quantum
dissipative systems \cite{weiss-book,leggett-etal}. It plays an 
important role in describing defect tunneling in solids
\cite{grabert-wipf,golding-etal}, quantum tunneling between flux states in a
SQUID \cite{leggett}, electron tunneling between quantum dots
\cite{fujisawa-etal}, or electron transfer in chemical and biological
reactions \cite{chandler}. The model provides a
nontrivial description of dissipation in a quantum system and has 
attracted much interest due to its simplicity and applicability.
It consists of a two-state (spin) system coupled linearly to an 
infinite-dimensional harmonic oscillator (boson) bath with the Hamiltonian
$H=H_0+H_B+V$, where
$H_0 = -\frac{\Delta}{2}\sigma_x + \frac{\epsilon}{2}\sigma_z$, 
$H_B = \sum_q \omega_qa^{\dagger}_qa_q$, and 
$V = \frac{\sigma_z}{2}\sum_q g_q (a^{\dagger}_q+a_q)$.
Here, $\sigma_x$ and $\sigma_z$ are the usual Pauli matrices. $\Delta$
is the tunnel matrix element and $\epsilon$ the energy difference
between the two states. $a^{\dagger}_q$ ($a_q$) creates (annihilates) 
a boson with energy $\hbar\omega_q$. The quantity $g_q$ represents the 
coupling strength of the two-state system to the coordinate of the $q^{th}$
oscillator. The coupling to the environment is completely defined by
the spectral density $J(\omega) = \pi\sum_q g_q^2 \delta(\omega-\omega_q)$,
which is usually parametrized by 
$J(\omega) = 2\pi\alpha\omega^{n+1}e^{-\omega/D}$.
The case $n=0$ corresponds to the Ohmic bath, which we want to consider in the
following. Therefore the coupling to the bath is now characterized by
the coupling strength $\alpha$ and the high-energy cutoff $D$.

Except for some special parameter values an exact solution of
the spin-boson model is not known \cite{sassetti-weiss-1}. 
Most of the studies are based on perturbative approaches in $\alpha$ 
\cite{slichter} or $\Delta$ \cite{leggett-etal}. The latter is known as
the noninteracting blip approximation (NIBA) and gives very reliable
results at zero bias for the diagonal matrix elements of the reduced
density matrix $p(t)$ of the two-level system. Recent real-time quantum 
Monte Carlo 
(QMC) simulations \cite{egger-mak,stockburger-mak} provide also reliable
information on the nondiagonal elements of $p(t)$, but they were only applied
to the symmetric case and the correct long-time behaviour has not been checked.
Flow equation methods based on infinitesimal
unitary transformations \cite{kehrein-mielke} have reproduced 
the Shiba-relation up to $\alpha\sim 0.025-0.05$ (with an error of 
$3-10\%$), but they only addressed spectral properties of the system.
Other approaches try to use a mapping of the spin-boson
model on the anisotropic Kondo model \cite{mapping}, and solve the 
latter exactly using numerical renormalization group (NRG) 
\cite{costi-1}, Bethe ansatz \cite{costi-zarand}, or conformal field 
theory (CFT) \cite{lesage-saleur}. However, NRG and Bethe ansatz provide only 
spectral properties or dynamics at very short time scales \cite{costi-2}, 
and CFT has solved so far only the unbiased case $\epsilon=0$
for the diagonal elements of $p(t)$. Furthermore,
and most importantly, the mapping on the Kondo model can not
be proven rigorously, and the relation of the parameters
is not precisely known \cite{leggett-etal,weinmann}. It is known  
that the mapping is incorrect for finite cutoff D, but it is at 
least established that the scaling behaviour agrees with that of the 
spin-boson model \cite{kehrein-mielke,costi-1,lesage-saleur}.

In this paper, we will present for the first time a solution of
the complete dynamics of the spin-boson model for $\alpha\lesssim 0.1-0.2$.
We study the diagonal and nondiagonal part of the reduced density 
matrix $p(t)$ starting from an arbitrary nonequilibrium
state $p_0$. From the asymptotic behaviour we determine the relaxation and 
dephasing time. Furthermore, we calculate the spin susceptibility
and correlation functions. Especially, and in contrast to 
many other methods, we solve directly the spin-boson model and
present results at finite cutoff $D$ and finite bias $\epsilon$.
Therefore, our results provide for the first time the possibility
for a quantitative and unambigious comparism to results obtained 
from the mapping of the spin-boson model on the Kondo model.
We find that for $\epsilon=0$ the relaxation
parameters agree with those of CFT and in the scaling limit the susceptibility
agrees rather well with Bethe-ansatz results, but for the latter
deviations occur at finite bias. To demonstrate the reliability 
of our results, we show the consistency with chromostochastic 
quantum dynamics (CSQD) \cite{stockburger-mak}, and check the 
Shiba-relation as well as the scaling behaviour. 

To obtain our results, we will use a recently developed real-time
renormalization group (RTRG) method \cite{koenig-hs,hs}. This method has 
been successfully applied
to equilibrium problems \cite{koenig-hs,mk-hs}, and to the 
study of nonequilibrium stationary states \cite{hs-koenig}.
Here we will apply it for the first time to the
dynamics of the reduced density matrix, and 
generalize it to the evaluation of correlation functions as well.
Since this technique is rather straightforward
and easily generalized to other models, this may open a 
new possibility for the study of various kinds of dissipative
quantum systems, like e.g. many-level systems, magnetic nanoparticles
interacting with phonons, coupled quantum dots or other kinds of 
dissipative environments.

{\it Kinetic equation and RTRG approach}.
The RTRG approach is based on a kinetic equation for the density
matrix, see Ref.~\cite{hs} for details. We only mention the main steps:

1. The time evolution of the reduced density matrix is written
in Liouville space as $p(t) = {\rm Tr}_B \exp{(-iLt)} p_0\rho_B^{eq}$, where
${\rm Tr}_B$ denotes the trace over the bath degrees of freedom, and
$L=[H,\cdot]=L_0+L_B+L_V$ is the Liouville operator. The initial 
density matrix is assumed to decouple into an arbitrary nonequilibrium 
distribution $p_0$ for the two-level system
and an equilibrium distribution $\rho_B^{eq}$ for the oscillator
bath. 

2. The propagator $\exp{(-iLt)}$ is expanded in the interaction 
part $L_V$, and the trace ${\rm Tr}_B$ over the bath degrees of freedom is 
performed by application of Wick's theorem. In this way one obtains a 
series of terms where vertices $G^p$ of the local system (originating
from the $\sigma_z/2$ factor in $V$) are connected
by pair contractions $\gamma^{pp'}(t)$ of the bath. Here, $p=\pm$ indicates
wether the interaction takes place on the forward or the backward 
propagator. With $j=\sum_q g_q (a^\dagger_q + a_q)$ and 
$\gamma(t) = Tr_B \rho_B^{eq} j(t) j$, we obtain 
$\gamma^{pp'}(t)=R(t)+ip'S(t)$ with $\gamma(t)=R(t)+iS(t)$ and
\begin{eqnarray}
\label{R}
R(t) &=& -2\alpha{\rm Re}\{(\pi
T)^2\frac{1}{\sinh^2(\pi T(t-i/D))}\}\,,\\
\label{S}
S(t) &=& -2\alpha{\rm Im}\{\frac{1}{(t-i/D)^2}\}\,,
\end{eqnarray}
where we restricted ourselves to the physically relevant situation
$D\gg T$. 

3. From the diagrammatic language one can derive a formally exact
kinetic equation for $p(t)$ 
\begin{equation}
\label{kineticeq}
\dot{p}(t) + iL_0p(t) = \int_0^tdt'\,\Sigma(t-t')p(t')\,,
\end{equation}
where $\Sigma(t-t')$ is a superoperator acting on $p(t')$ and
is defined by the sum over all irreducible diagrams. In Laplace space 
we get the formal solution $p(z) = \Pi(z)p_0$ with
$\Pi(z) = i/(z-L_0-i\Sigma(z))$.

4. The kernel $\Sigma(z)$ is calculated by a renormalization group  
procedure. Short time scales of $\gamma(t)$ are integrated out first 
by introducing a short-time cutoff $t_c$ into the correlation
function $\gamma(t)\rightarrow\gamma(t,t_c)$. In each renormalization
group step, the time scales between $t_c$ and $t_c+dt_c$ are 
integrated out, starting from $t_c=0$ and ending at $t_c=\infty$.
As a consequence, one generates RG equations for $\Sigma(z)$, 
$L_0$, $G^p$, and the two boundary vertex operators $A^p$ and $B^p$ 
(defined as the rightmost and leftmost vertex of the kernel $\Sigma(z)$).
Within the scheme of a perturbative RG analysis, generation of multiple
vertex-operators is neglected here, as in Ref.~\cite{hs-koenig}.
For the present model, we choose the cutoff dependence either as
$\gamma(t,t_c)=\gamma(t)\Theta(t-t_c)$ (choice I) or
\begin{equation}
\label{newdef}
\gamma(t,t_c) = \frac{d}{dt}\left(\tilde{R}(t)\Theta(t-t_c)\right) +
iS(t)\Theta(t-t_c)\,,
\end{equation}
with $R(t) = (d/dt) \tilde{R}(t)$ (choice II). It turns out that
choice I is better for $t_c \gtrsim \Delta$, whereas choice II
is needed for small $t_c$ in order to avoid unphysical
linear dependencies on the cutoff $D$. At $t_c^0$ we use
a smooth crossover from choice I to choice II. Solving the RG equations
numerically gives the kernel $\Sigma(z)$ in Laplace
space, and the reduced density matrix can be deduced. The static 
susceptibility follows from
$\chi_0=-(d/d\epsilon) {\rm Tr_0} \sigma_z p_{st}$, where ${\rm Tr_0}$ 
is the trace over the local system, and $p_{st}$ denotes the
stationary solution $p_{st}=\lim_{t\rightarrow\infty}p(t)$.

5. To calculate correlation functions of the form $\chi(t) =
\frac{1}{2}{\rm Tr}\{[\sigma_z(t),\sigma_z]\rho^{eq}\}$ we have to
generalize the RG procedure. With $C = G^+ + G^-$ we get in Laplace
space
\begin{equation}
\chi(z) = {\rm Tr_0}\{\sigma_z\Pi(z)(C+\Sigma_C(z))p_{st}\}\,,
\end{equation}
where $\Sigma_C(z)$ is analogously defined to $\Sigma(z)$ but
contains the ``vertex'' $C$ at any time point. Within the
framework of Ref.~\cite{hs}, the RG equation for $\Sigma_C(z)$
can be easily derived and reads
\begin{eqnarray}
{d\over dt_c}\Sigma_C(z)\,&=&\,\sum_{pp'}\int_0^\infty dt \int_0^t dt' 
\,\frac{d}{dt_c}\gamma^{pp'}(t,t_c)\nonumber\\
&&e^{iz(t-t')}A^p e^{-iL_0(t-t')}C 
e^{-iL_0 t'}\bar{B}^{p'}\,,
\end{eqnarray}
where $\bar{B}^p= B^p|_{z=0}$.

{\it Results}.
Figs.~\ref{dm1} and \ref{dm2} show the time evolution of $p(t)$
for the unbiased and biased case. Initially, the two-level system 
is prepared in the spin-up state ($u$). For
$\epsilon=0$, we achieve a very good agreement with CSQD. 
Only the real parts of the nondiagonal elements, which correspond 
to $\langle\sigma_x\rangle$, exhibit a deviation of approximately 
$5\%$. However, the CSQD can not give an accurate
error for $\langle\sigma_x\rangle$ \cite{stockburger-pc}.  
The diagonal elements oscillate in time with the frequency 
$\Delta_r\sim \Delta(\Delta/D)^{\alpha/(1-\alpha)}$. This agrees with
the renormalized tunnel matrix element, which is the characteristic
energy scale of the spin boson model \cite{weiss-book}. In Fig.~\ref{dm1}
we obtain $\Delta_r=0.626\Delta$ (CFT gives $\Delta_r=0.625\Delta$). 
In contrast, the real part of the
nondiagonal elements is purely decaying for $\epsilon=0$ (only for
$\epsilon\ne 0$ it also exhibits oscillations). In the scaling
limit, i.e. for $D,\Delta\rightarrow\infty$ such that $\Delta_r={\rm const}$,
the diagonal elements are universal, i.e. they only depend on
$\Delta_r t$. The nondiagonal elements however have an extra factor of
$\Delta_r/\Delta$ (see inset in Fig.~\ref{dm1}) which is consistent with
\cite{weiss-book}. The long-time
behaviour is given by an exponential decay. For $\epsilon=0$ the decay
constants of the diagonal (nondiagonal) elements correspond to the relaxation
(dephasing) time $\tau^{rel}$ ($\tau^{dep}$), see insets in 
Fig.~\ref{dm2}. For $\tau^{rel}$ we find good agreement
with CFT.For
$\epsilon\neq 0$ the diagonal (nondiagonal)
elements also contain an incoherent (coherent) decay with $\tau^{dep}$
($\tau^{rel}$). We note that our zero-temperature results at
finite bias and the dephasing time are the first ones presented in the
literature.

The static susceptibility for the unbiased and biased case is shown 
in Figs.~\ref{sus1} and \ref{sus2}. Clearly, the susceptibility
depends strongly on the cutoff $D$ (see right inset in Fig.~\ref{sus1}). 
However, in the scaling limit the susceptibility is universal. We normalize
our results using the zero-temperature value $\chi_0(T=0)$.
For $\epsilon=0$, the susceptibility is a 
monotonous function, whereas a local maximum is obtained 
for the biased case. A quantitative comparism with recent Bethe 
ansatz results for the Kondo model shows a very good agreement for
the unbiased case (see left inset in Fig.~\ref{sus1}). 
However, for $\epsilon\ne 0$,
there are deviations which demonstrates that the commonly used relation 
of the parameters of the Kondo model to the ones of the spin-boson model 
has to be taken with care. In the high-temperature limit we obtain
the correct $(1/2T)$ law, which is independent of $\alpha$, $\epsilon$ and $D$
\cite{costi-zarand}.

The result for the imaginary part of the dynamic susceptibility
$\chi''(\omega)=\int_{-\infty}^{\infty}d\omega\,e^{i\omega
t}\frac{1}{2}{\rm Tr}\{[\sigma_z(t),\sigma_z]\rho^{eq}\}$ is shown in
Fig.~\ref{chi}, with $S(\omega)=\chi''(\omega)/\omega$.
The Shiba-relation provides an exact equation for the low frequency 
limit of $\chi''$ at $T=0$ \cite{shiba}
\begin{equation}
\label{shibaeq}
\lim_{\omega\rightarrow 0}S(\omega)=2\pi\alpha\chi_0^2\,.
\end{equation}
We tested Eq.~(\ref{shibaeq}) for different $\alpha$, $\epsilon$ and $D$, 
see Tab. \ref{shibatab}. For $\alpha\lesssim 0.1$, we achieved 
a very good agreement (error smaller than $5\%$).
Like the static susceptibility, the correlation function depends
strongly on the cutoff $D$, but in the scaling
limit, the normalized 
quantity $S(\omega)/S(0)$ depends only on $\omega/\omega_{max}$ (see inset in
Fig.~\ref{chi}), where $\omega_{max}\sim\Delta_r$ 
denotes the frequency where $S(\omega)$ is maximum.
We note that
our results for the correlation function are the first ones presented
for the spin-boson model for $\alpha=0.1$. NRG results \cite{costi-1}
are very accurate for low frequency but fail for $\omega\sim\Delta$
\cite{kehrein-mielke}, flow equation methods have already an error 
of $\sim 25\%$ concerning the Shiba-relation \cite{kehrein-mielke}, 
and CSQD does not provide a check of the Shiba-relation and only gives
data for the unbiased case $\epsilon=0$ 
\cite{stockburger-mak}.

In summary, we investigated the spin-boson model for ohmic dissipation
using real-time renormalization group. For coupling parameters
$\alpha\lesssim 0.1-0.2$ we achieved a full solution of the static
and dynamical properties. We calculated the real-time evolution
of all matrix elements of the reduced density matrix together with
the static susceptibility and correlation functions. In contrast to
previous works we are not restricted to zero bias or the scaling
limit. The restriction in $\alpha$ is due to the fact that we neglected
double vertex diagrams. Our results show that the RTRG method is a
very flexible tool to treat various kinds of dissipative quantum
systems.

{\it Acknowledgments}.
We acknowledge useful discussions with K. Sch\"onhammer and
U. Weiss. We also thank J. T. Stockburger and T. A. Costi for their
numerical data. This work was supported by the ''Deutsche
Forschungsgemeinschaft'' as part of ''SFB 345'' (M.K.) and ``SFB 195'' (H.S.).

\begin{figure}
\centerline{\psfig{figure=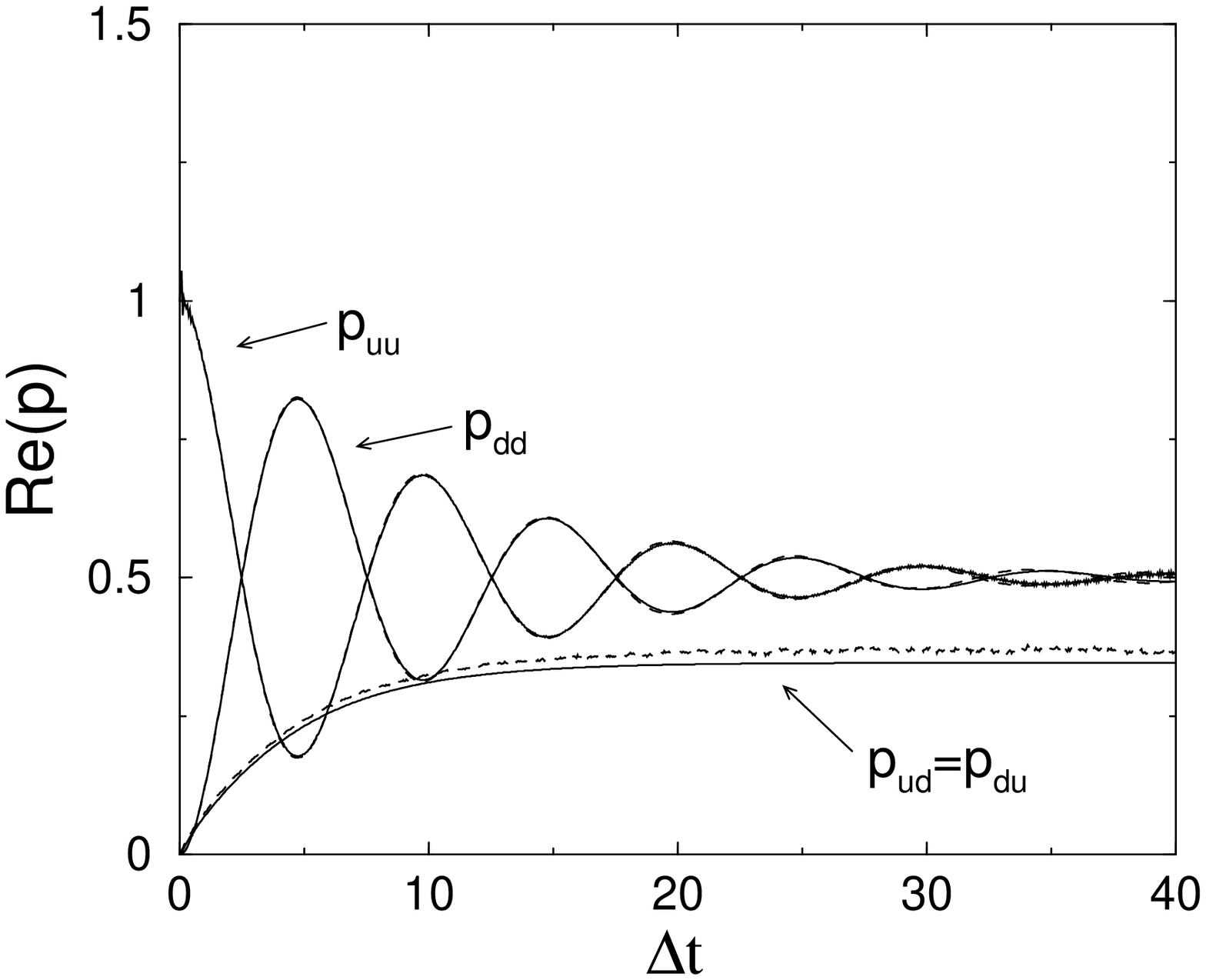,width=8cm,height=5cm,angle=0}}
\begin{picture}(0,0)
\put(117.0,85.0)
{
\begin{picture}(0,0)
\psfig{figure=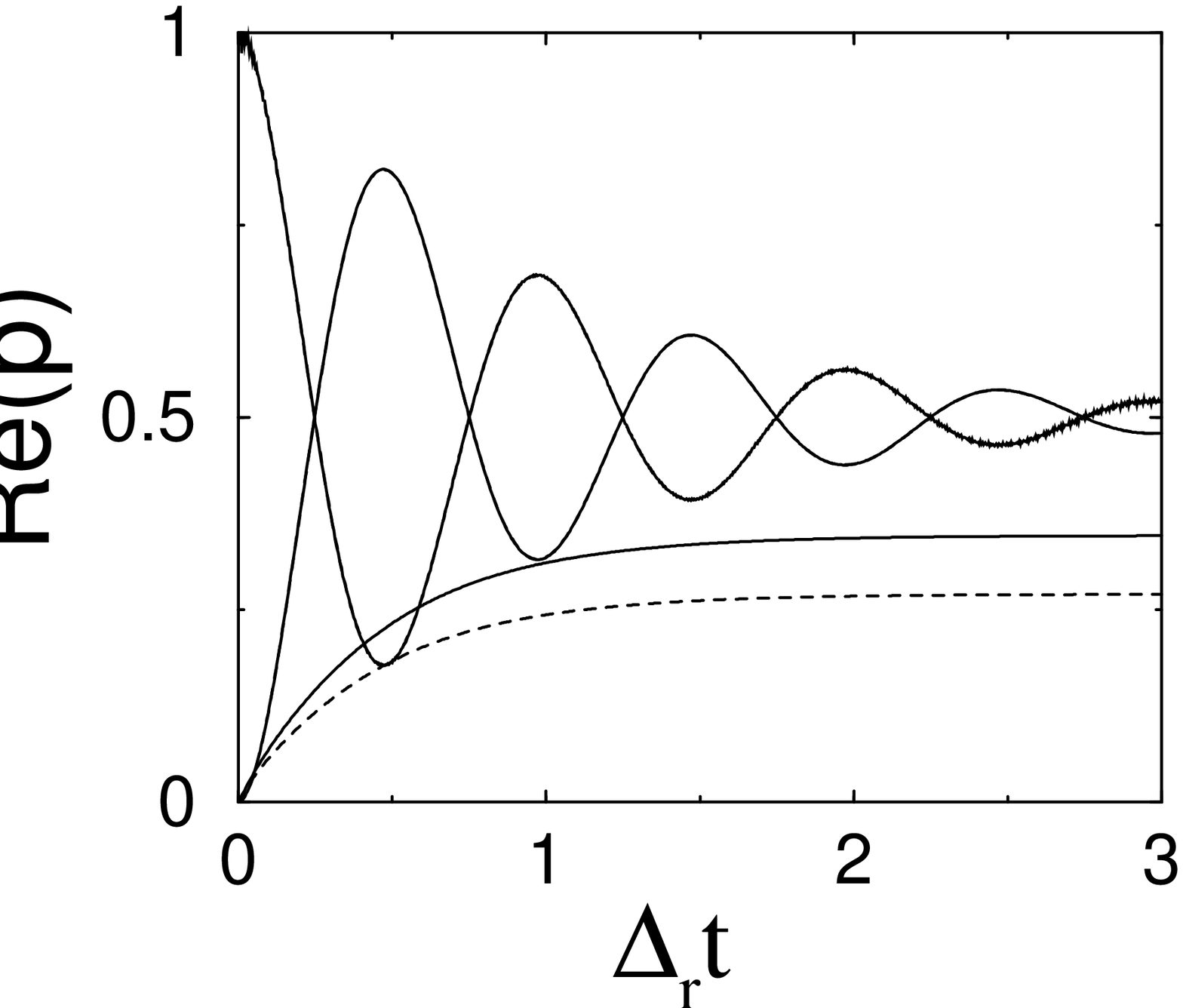,width=3.5cm,height=2.2cm,angle=0}
\end{picture}
}
\end{picture}
\caption{Real part of $p(t)$ for $\alpha=0.1$, $\epsilon=0$,
  $D=100\Delta$ and $T=0$. Solid lines: RTRG.
  Dashed lines: CSQD; Inset: Rescaled real part of $p(t)$. 
  Solid lines: $D=100\Delta$. Dashed lines: $D=1000\Delta$.}
\label{dm1}
\end{figure}

\begin{figure}
\centerline{\psfig{figure=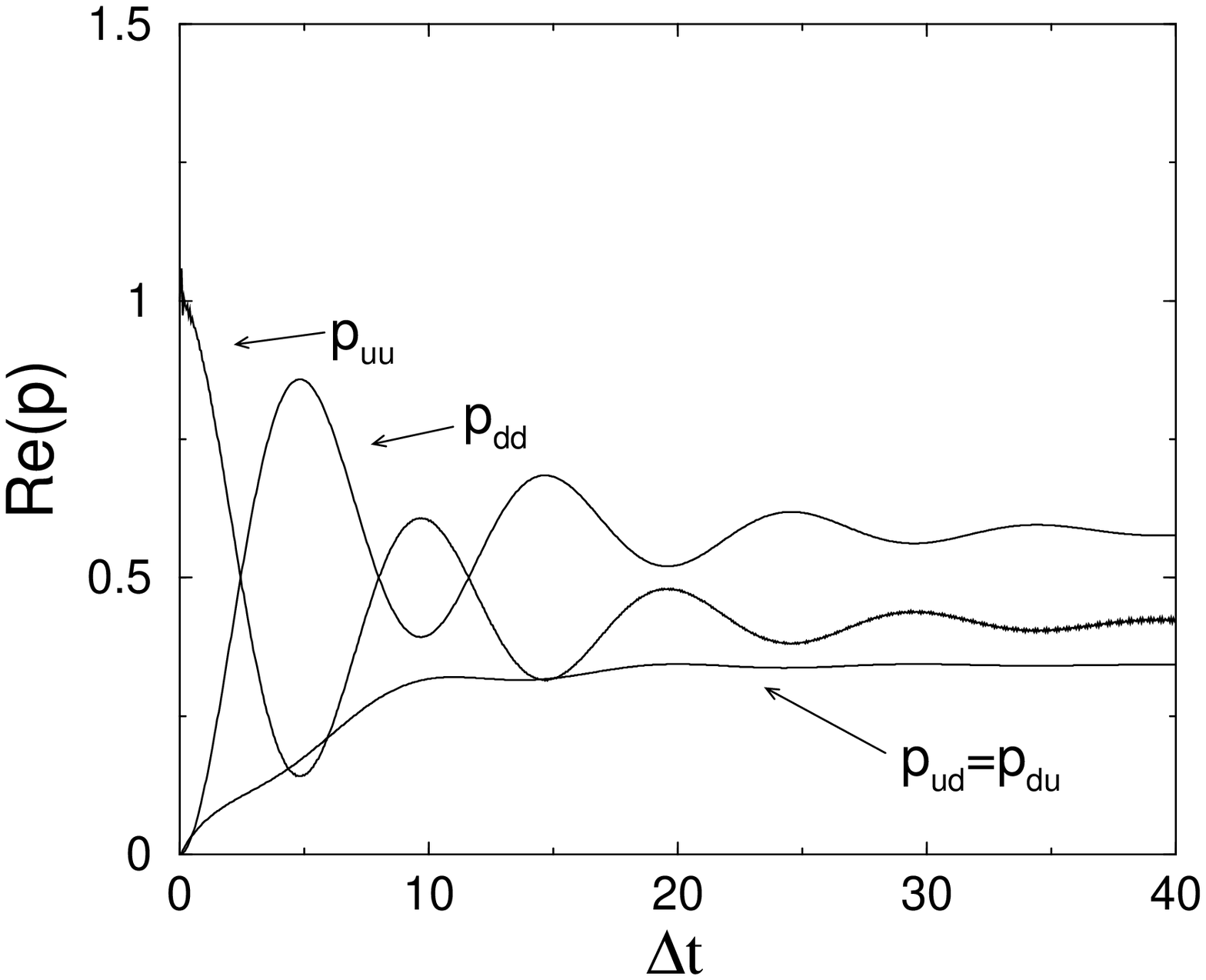,width=8cm,height=5cm,angle=0}}
\begin{picture}(0,0)
\put(66.0,100.0)
{
\begin{picture}(0,0)
\psfig{figure=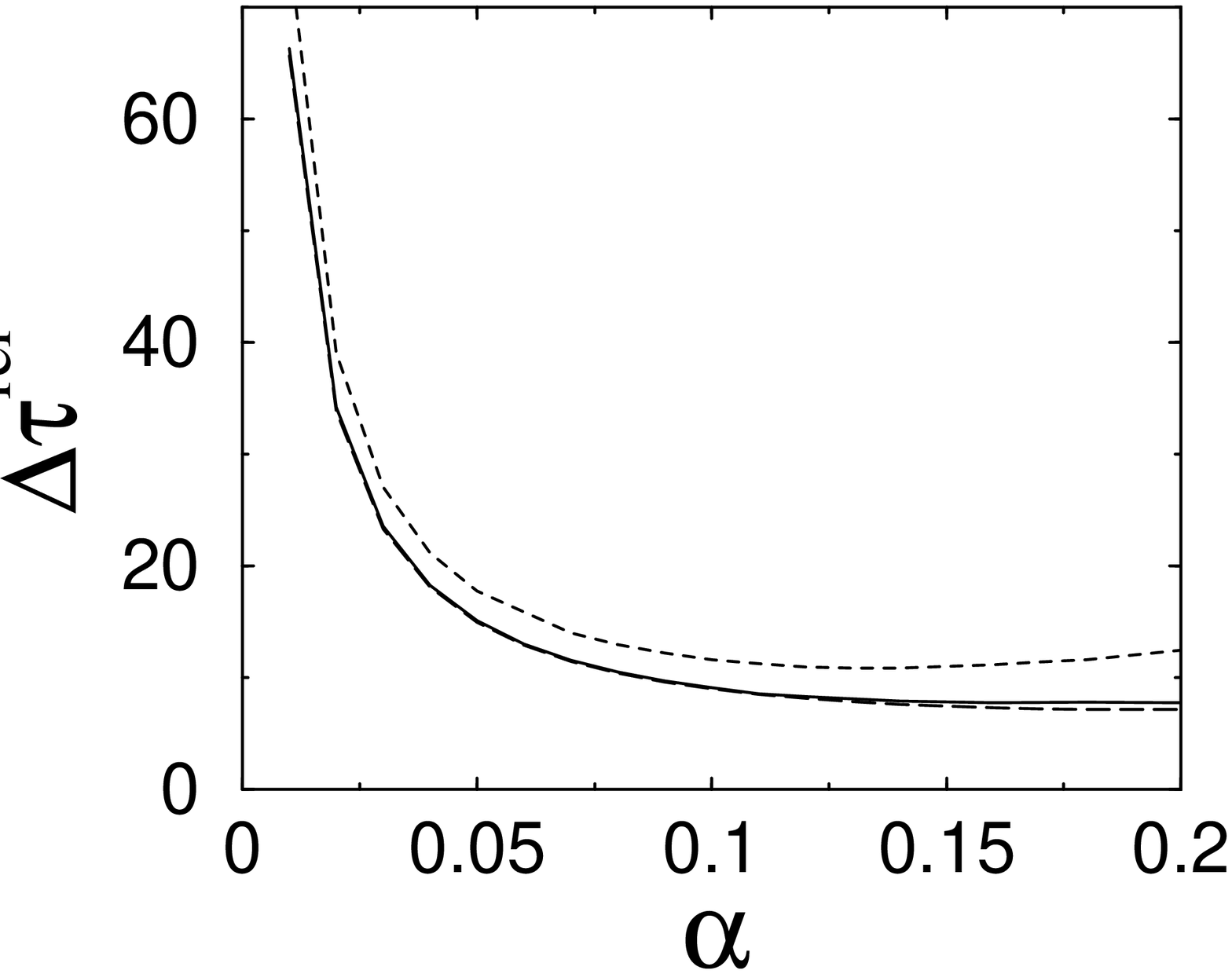,width=2.6cm,height=1.6cm,angle=0}
\end{picture}
}
\end{picture}
\begin{picture}(0,0)
\put(140.0,100.0)
{
\begin{picture}(0,0)
\psfig{figure=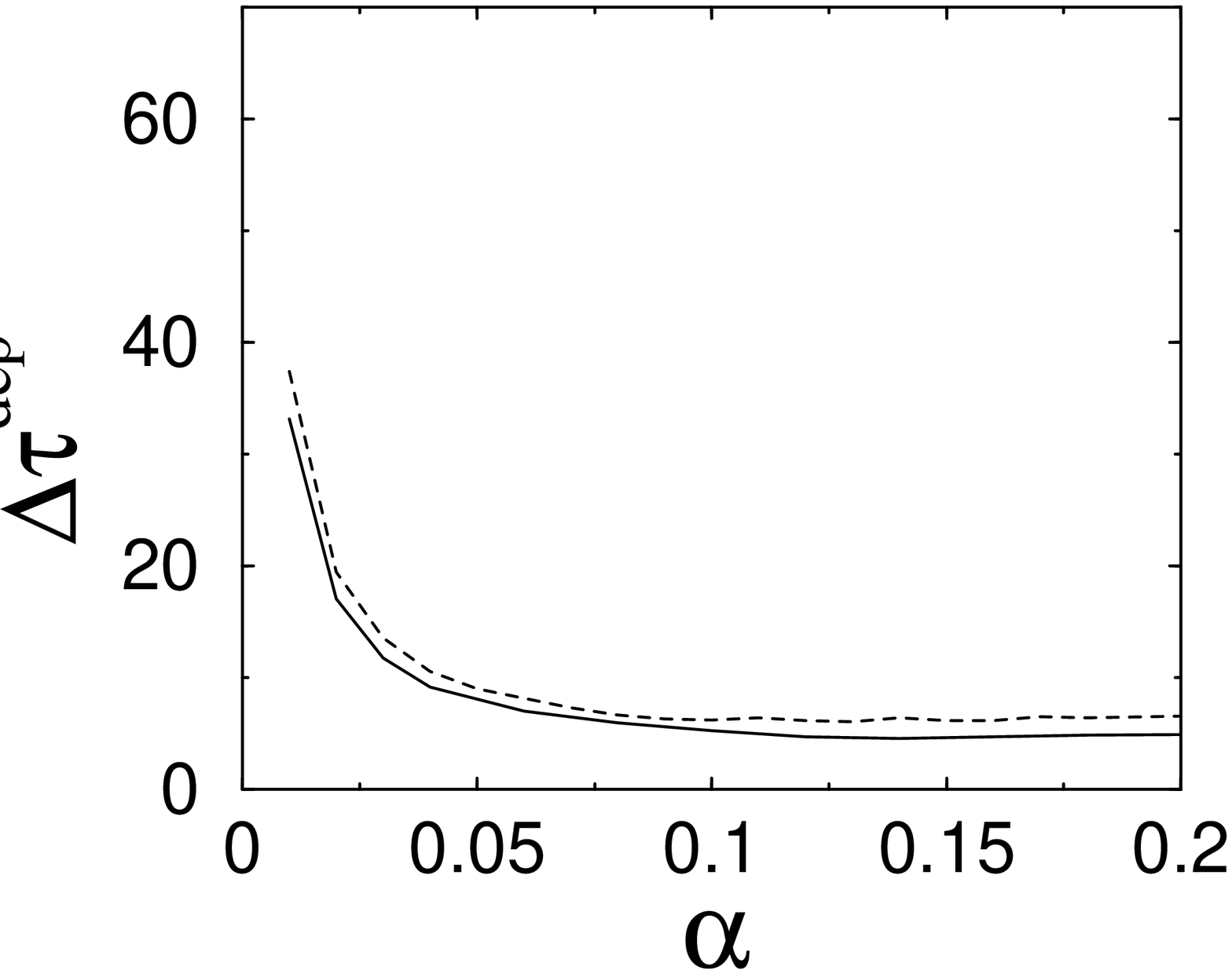,width=2.6cm,height=1.6cm,angle=0}
\end{picture}
}
\end{picture}
\caption{Real part of $p(t)$ for $\alpha=0.1$, $\epsilon=0.1\Delta$,
  $D=100\Delta$ and $T=0$;
  Left inset: $\tau^{rel}$ as a function of $\alpha$ for $D=100\Delta$ and
  $T=0$. Solid line: $\epsilon=0$. Dashed line:
  $\epsilon=0.5\Delta$. Long dashed line: CFT, $\epsilon=0$; Right
  inset: $\tau^{dep}$ as a function of $\alpha$ for $D=100\Delta$ and
  $T=0$. Solid line: $\epsilon=0$. Dashed line: $\epsilon=0.5\Delta$.} 
\label{dm2}
\end{figure}

\begin{figure}
\centerline{\psfig{figure=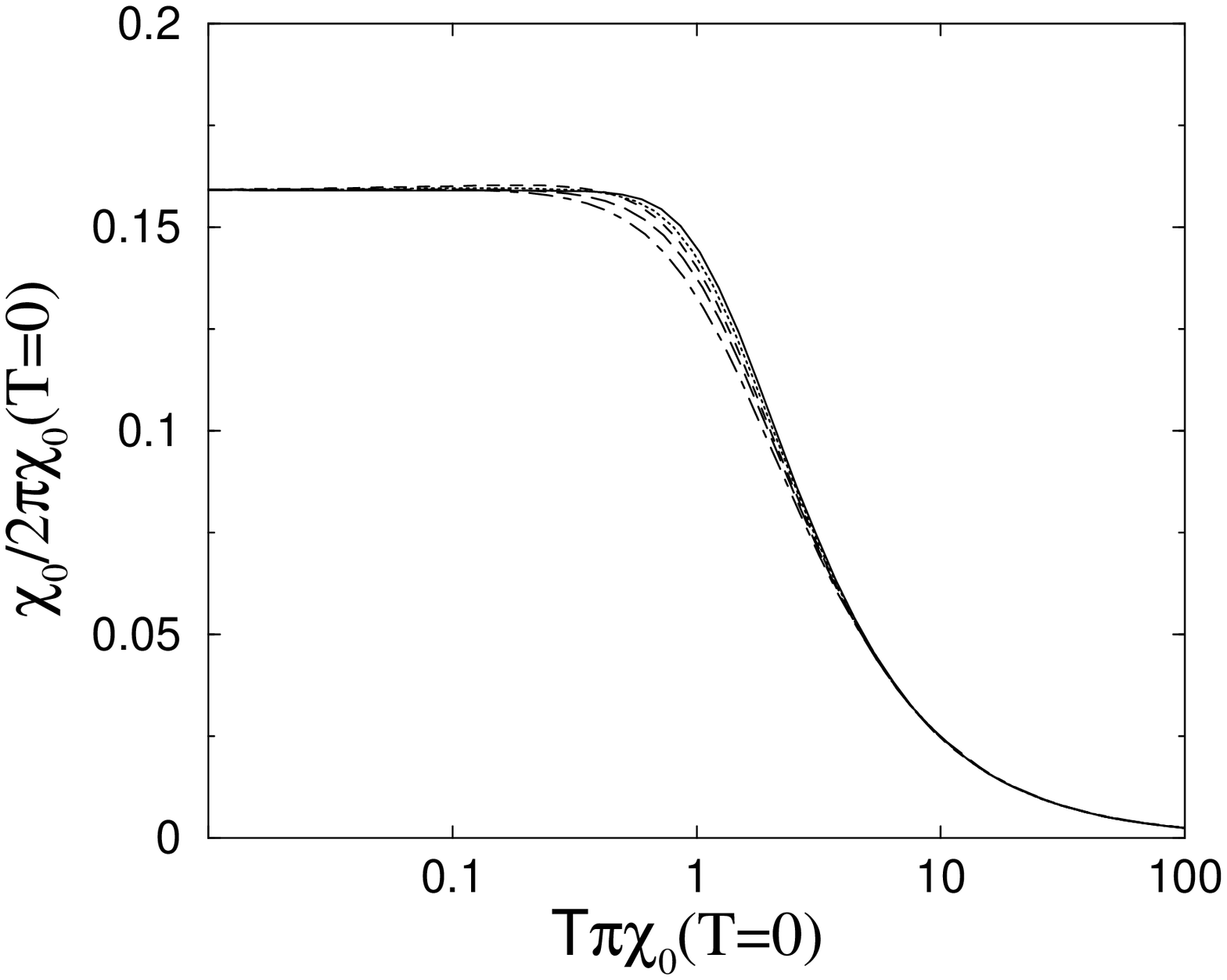,width=8cm,height=5cm,angle=0}}
\begin{picture}(0,0)
\put(47.0,45.0)
{
\begin{picture}(0,0)
\psfig{figure=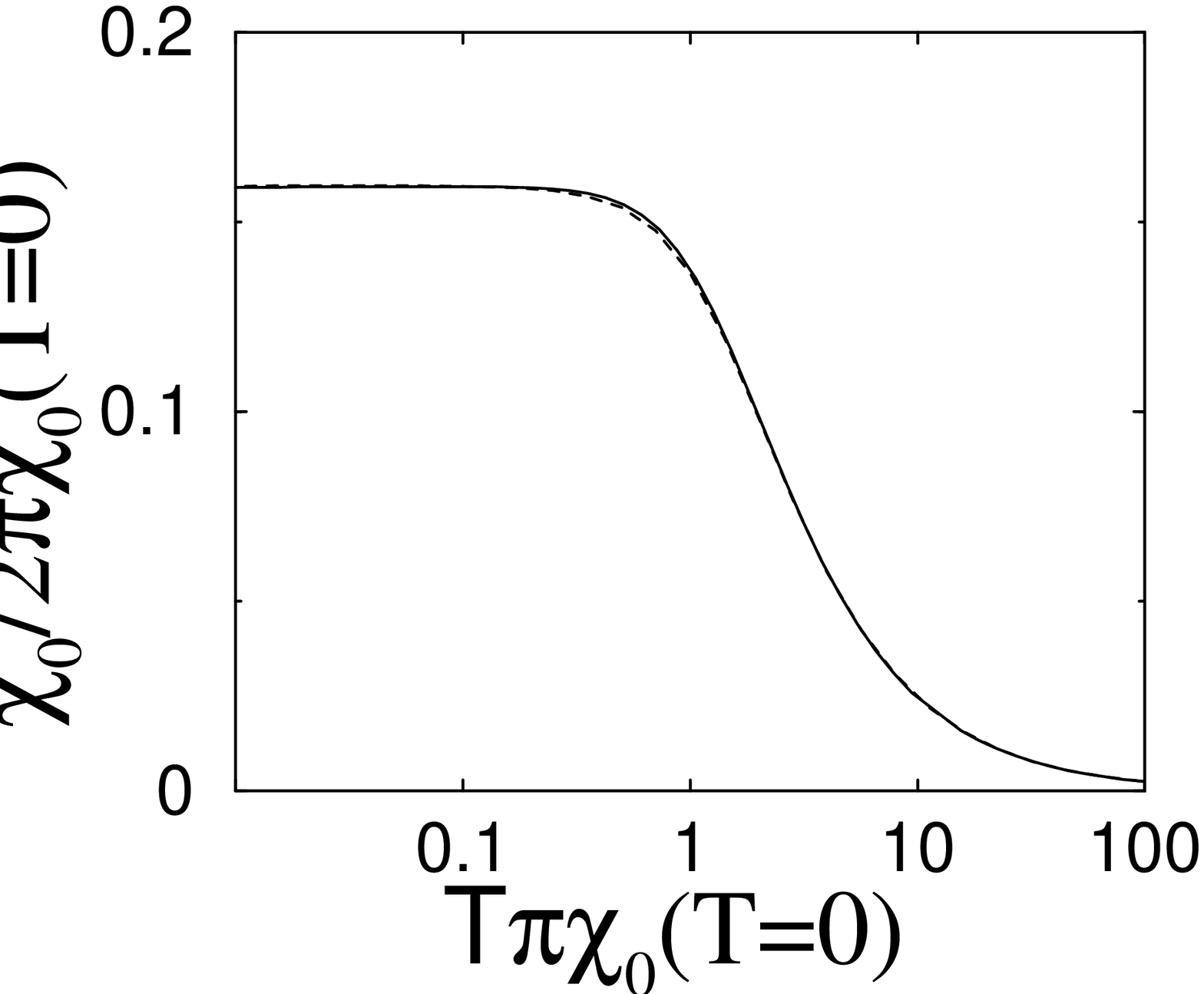,width=3.0cm,height=1.9cm,angle=0}
\end{picture}
}
\end{picture}
\begin{picture}(0,0)
\put(133.0,92.0)
{
\begin{picture}(0,0)
\psfig{figure=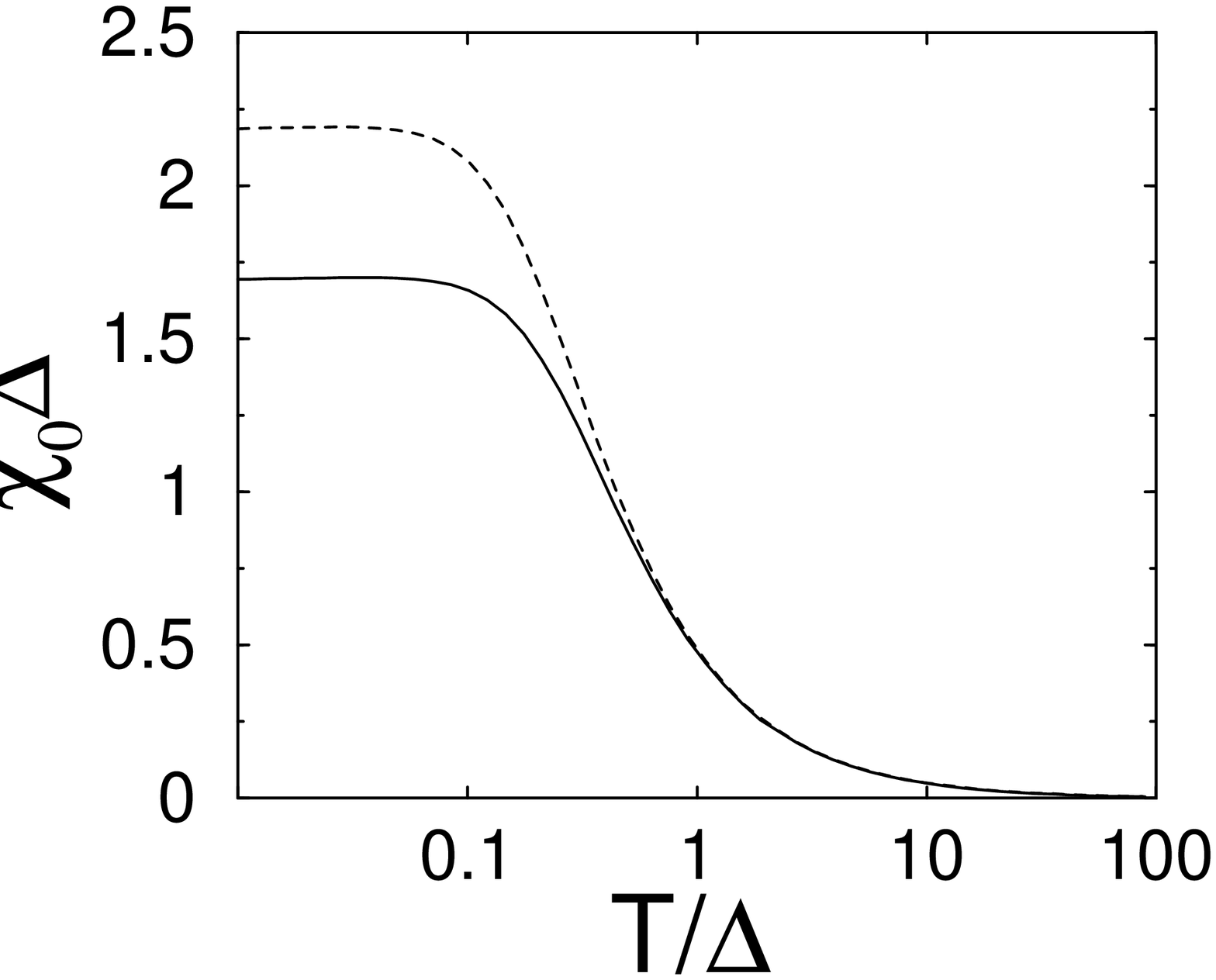,width=2.8cm,height=2.0cm,angle=0}
\end{picture}
}
\end{picture}
\caption{Static susceptibility as a function of temperature for
  $\epsilon=0$, $D\gg\Delta$, and $\alpha=0.01,0.05,0.1,0.125,0.2$
  (from top to bottom);
  Left inset: $\alpha=0.125$. Solid line: RTRG. Dashed line: Bethe ansatz;
  Right inset: Cutoff dependence for $\alpha=0.1$. Solid line:
  $D=100\Delta$. Dashed line: $D=1000\Delta$.}
\label{sus1}
\end{figure}

\begin{figure}
\centerline{\psfig{figure=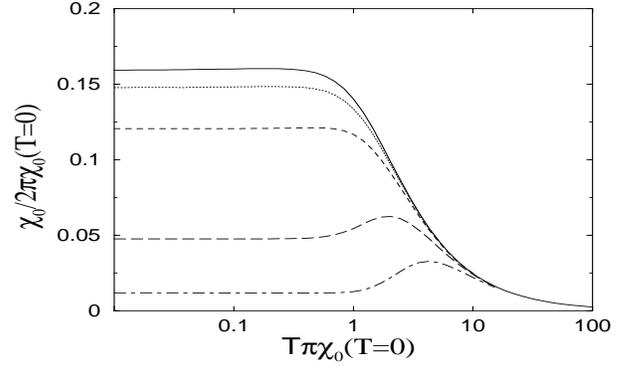,width=8cm,height=5.0cm,angle=0}}
\caption{Static susceptibility as a function of temperature for
  $\alpha=0.1$, $D\gg\Delta$, and $\epsilon/\Delta=0,0.1,0.2,0.5,1$
  (from top to bottom).}
\label{sus2}
\end{figure}

\begin{figure}
\centerline{\psfig{figure=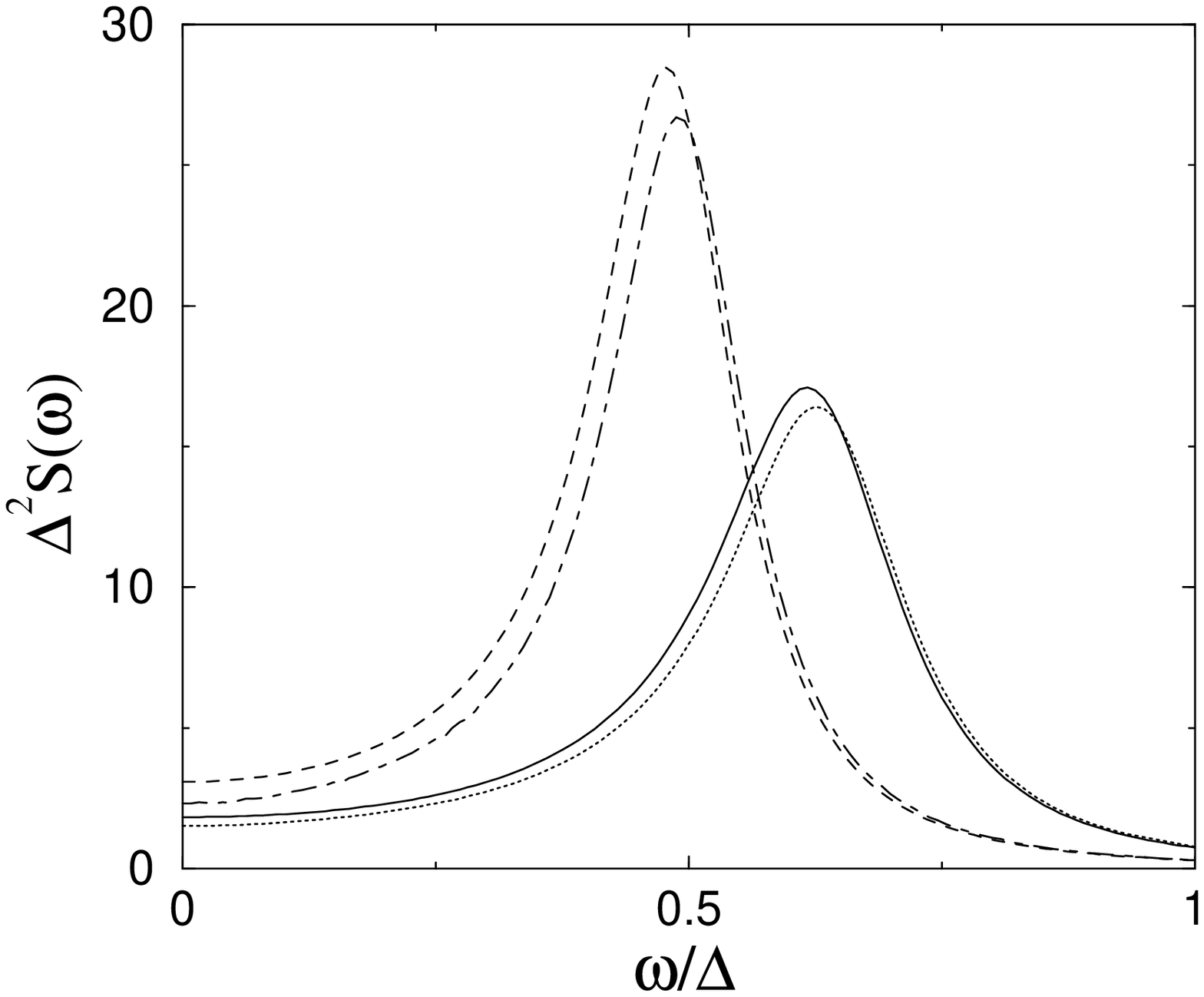,width=8cm,height=5cm,angle=0}}
\begin{picture}(0,0)
\put(143.0,93.0)
{
\begin{picture}(0,0)
\psfig{figure=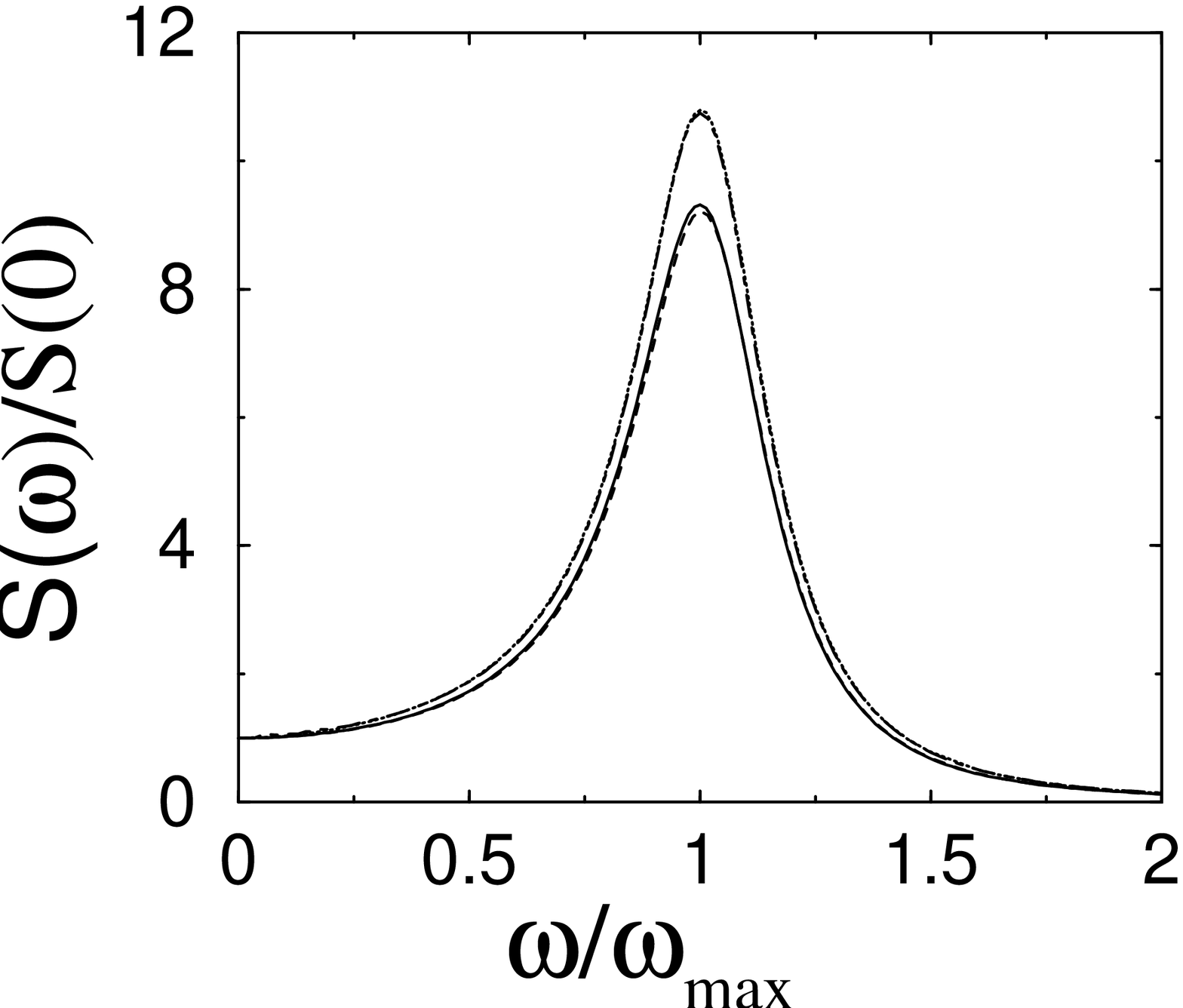,width=2.7cm,height=2.0cm,angle=0}
\end{picture}
}
\end{picture}
\caption{$S$ as a function of $\omega$ for $\alpha=0.1$,
  $T=0$. Solid line: $\epsilon=0$,
  $D=100\Delta$. Dotted line: $\epsilon=0.1\Delta$, $D=100\Delta$. Dashed
  line: $\epsilon=0$, $D=1000\Delta$. Dot-dashed
  line: $\epsilon=0.1\Delta$, $D=1000\Delta$;
  Inset: Rescaled $S$. Upper curve: $\epsilon=0$. 
  Lower curve: $\epsilon=0.1\Delta$ (for $D=1000\Delta$ we
  rescaled $\epsilon=(1000/100)^{\alpha/(1-\alpha)}\times 0.1\Delta
  =10^{1/9}\times 0.1\Delta$).}
\label{chi}
\end{figure}

\begin{table}
\begin{tabular}{ddrddd}
$\alpha$ & $\epsilon/\Delta$ & $D/\Delta$ & $\chi_0\Delta$ & $\lim\limits_{\omega\rightarrow 0}\Delta^2S(\omega)$ & error \\ \hline
0.01 & 0.0 & 100 & 1.0511 & 0.0680 & 2.04\% \\ 
0.05 & 0.0 & 100 & 1.2899 & 0.5083 & 2.80\% \\ 
0.1 & 0.0 & 100 & 1.6868 & 1.8343 & 2.57\% \\ 
0.2 & 0.0 & 100 & 3.2400 & 12.3085 & 6.93\% \\ 
0.01 & 0.0 & 1000 & 1.0772 & 0.0711 & 2.50\% \\ 
0.05 & 0.0 & 1000 & 1.4534 & 0.6445 & 2.92\% \\ 
0.1 & 0.0 & 1000 & 2.1804 & 3.0880 & 3.33\% \\ 
0.2 & 0.0 & 1000 & 5.7219 & 35.4919 & 14.75\% \\ 
0.1 & 0.01 & 100 & 1.6859 & 1.8309 & 2.49\% \\
0.1 & 0.05 & 100 & 1.6671 & 1.7523 & 0.34\% \\
0.1 & 0.1 & 100 & 1.6036 & 1.6209 & 0.32\% \\
0.1 & 0.01 & 1000 & 2.1778 & 3.0911 & 3.66\% \\
0.1 & 0.05 & 1000 & 2.1369 & 2.7233 & 5.21\% \\
0.1 & 0.1 & 1000 & 2.0187 & 2.4377 & 4.91\% \\
\end{tabular}
\caption{Shiba-relation for different $\alpha$, $\epsilon$ and $D$.}
\label{shibatab}
\end{table}

\end{document}